\documentclass[aps,prl,reprint,groupedaddress,showpacs]{revtex4-1}

\usepackage{amsmath}
\usepackage{wasysym}
\usepackage{graphicx, color}
\usepackage{hyperref}
\usepackage{ifpdf}
\newif\ifdebug
\debugtrue
\ifdebug
  \usepackage{soul}
  
  \newcommand{\del}[1]{\textcolor{blue}{\st{#1}}}
  \newcommand{\remark}[1]{\textcolor{blue}{#1}}
\else
  
  \newcommand{\del}[1]{}
  \newcommand{\remark}[1]{}
\fi

\begin{document}

\title{A Gravitational Wave Detector with Cosmological Reach}

\author{Sheila Dwyer}
\email[]{dwyer\_s@ligo-wa.caltech.edu}
\author{Daniel Sigg}
\affiliation{LIGO Hanford Observatory, PO Box 159, Richland, WA 99352, USA}
\author{Stefan W. Ballmer}
\email[]{sballmer@syr.edu}
\affiliation{Department of Physics, Syracuse University, NY 13244, USA}
\author{Lisa Barsotti}
\author{Nergis Mavalvala}
\author{Matthew Evans}
\affiliation{Massachusetts Institute of Technology, Cambridge, MA 02139, USA}

\date{\today}

\begin{abstract}
Twenty years ago, construction began on the Laser Interferometer Gravitational-wave Observatory (LIGO). Advanced LIGO, with a factor of ten better design sensitivity than Initial LIGO, will begin taking data this year, and should soon make detections a monthly occurrence. While Advanced LIGO promises to make first detections of gravitational waves from the nearby universe, an additional factor of ten increase in sensitivity would put exciting science targets within reach by providing observations of binary black hole inspirals throughout most of the history of star formation, and high signal to noise observations of nearby events. Design studies for future detectors to date rely on significant technological advances that are futuristic and risky. In this paper we propose a different direction. We resurrect the idea of a using longer arm lengths coupled with largely proven technologies. Since the major noise sources that limit gravitational wave detectors do not scale trivially with the length of the detector, we study their impact and find that 40~km arm lengths are nearly optimal, and can incorporate currently available technologies to detect gravitational wave sources at cosmological distances $(z \gtrsim 7)$.
\end{abstract}

\pacs{04.80.Nn, 95.55.Ym, 95.85.Sz, 07.60.Ly}

\hypersetup{pdftitle={A Gravitational Wave Detector with Cosmological Reach}}
\hypersetup{pdfauthor={Sheila Dwyer, Daniel Sigg, Stefan W. Ballmer, Lisa Barsotti, Nergis Mavalvala, and Matt Evans}}
\hypersetup{pdfsubject={40 km Advanced LIGO}}
\hypersetup{pdfkeywords={interferometer gravitational-wave LIGO}}

\maketitle

\section{Introduction}
\label{sec:Introduction}

The current generation of gravitational wave detectors uses variants of long Michelson interferometers to detect minute deformations of space-time that pass through the Earth from distant astrophysical sources~\cite{weiss:1972, drever:1991,meers:1988}. Advanced LIGO~\cite{Fritschel2014b} employs Fabry-Perot arm cavities with a length of 4~km,  whereas Advanced VIRGO~\cite{degallaix:2012} and KAGRA~\cite{somiya:2011} are 3~km long. These instruments are likely to make direct detections of gravitational waves in the next several years~\cite{nsns}.  Coalescing neutron star binaries are expected to be a regular source for this generation of detectors, with sources at the horizon as far as 400 Mpc away. Observations of signals from pulsars, supernovae, and other sources are not ruled out, though they are likely to be infrequent and with low signal to noise ratios~\cite{nsns}.

\begin{figure}[t!]
\begin{center}
\centering\includegraphics[width = \columnwidth]{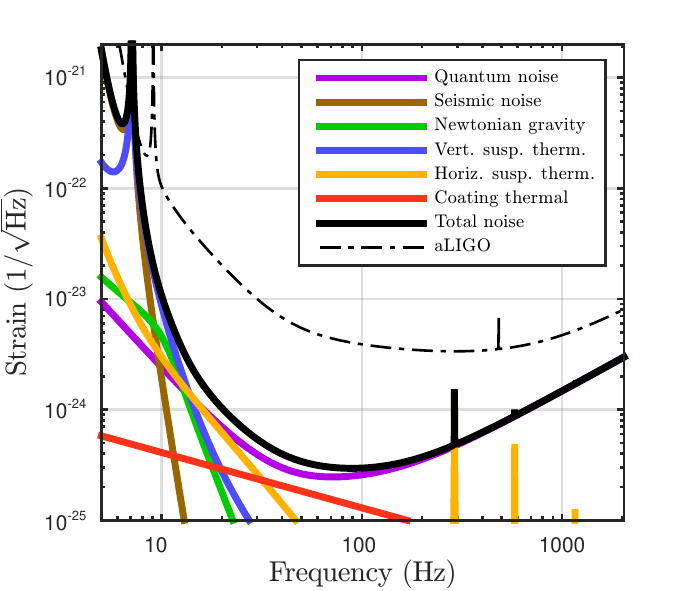}
\caption[Advanced LIGO sensitivity with extended arms.]{Projected sensitivity of a 40~km long interferometer based on Advanced LIGO. The only major added technology with respect to the existing interferometers is the use of a squeezed light source for reducing quantum noise.}
\label{fig:sens}
\end{center}
\end{figure}

Even as the scientific community prepares to gain new understanding of the nearby universe from the first detections of gravitational waves, the quest for deeper searches out to cosmological distances is a strong driving force toward significantly more sensitive detectors.  The reach of ground based detectors is limited by a class of noises known as displacement noises, which move the optics of the interferometer, and are to be contrasted with sensing noises, which limit the measurement of their position. Reducing displacement noises has been a major component of proposed upgrades to the current generation of detectors; a factor of two improvement in sensitivity is achievable through short-term incremental upgrades to Advanced LIGO \cite{Miller2014}. Later upgrades involving new optical materials and coatings, cryogenic operations, and other technologies currently being developed may achieve up to a factor of five improvement over Advanced LIGO in the existing 4 km facility \cite{adhikari:2014}.  Over time, increasingly complex upgrades in the existing facilities will yield smaller improvements in sensitivity.

To date the European Einstein Telescope proposal represents the most complete design of a future gravitational wave detector unfettered by existing facilities~\cite{abernathy:2011b}. The Einstein Telescope is 10~km long, underground, triangular shaped and has a projected astrophysical reach similar to the detector described in this paper, based on admittedly optimistic assumptions about improvements in technologies to reduce displacement noises.

We propose a much simpler approach to improving the sensitivity based on proven technologies: increasing the arm length of existing detectors from 4~km to 40~km.  This does not automatically guarantee a ten-fold increase in sensitivity, since all noise sources do not scale trivially with arm length. This approach has two significant advantages: in the early phases it will open up cosmological distances to direct observation with gravitational waves using technology already proven in second generation detectors, and it will provide a facility where even more sensitive detectors can be installed in the future by incorporating advanced technologies.

\begin{figure}[t!]
\centering\includegraphics[width = 3.4in]{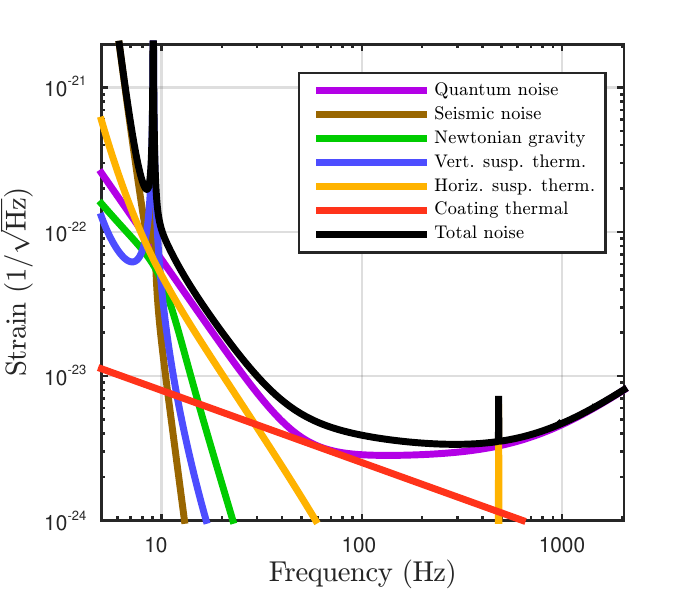}
\caption{The design noise budget of Advanced LIGO. All dominant noise sources below about 100~Hz are displacement noise, and therefore benefit from longer arms.}
\label{fig:aLIGO}
\end{figure}

This paper explores the sensitivity of a 40~km detector which, aside from arm length, requires only a few modest changes relative to the Advanced LIGO design (discussed in the second half of the paper). The projected sensitivity of this detector is shown in Figure~\ref{fig:sens}. Compared to Figure~\ref{fig:aLIGO} we see that it is possible to achieve an order of magnitude  improvement beyond Advanced LIGO, and also to move the most sensitive part of the detection band to lower frequencies where many astrophysical sources produce stronger signals. We go on to discuss the constraints on detector size which make the 40~km scale of particular interest, and why longer detectors move beyond the point of diminishing returns.

\section{Cosmological Reach}
\label{sec:Horizon}

\begin{figure}[t!]
\begin{center}
\includegraphics[width = 3.4in]{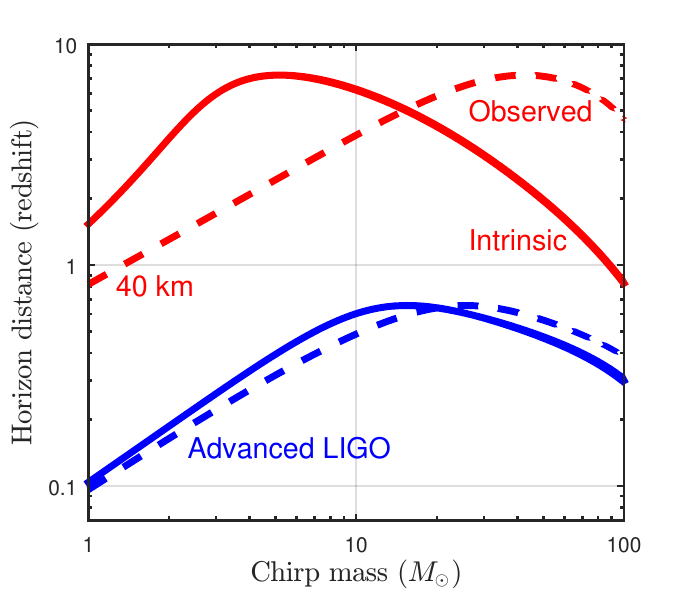}
\caption[Reach of Advanced LIGO with extended arms.]{Astrophysical reach for compact binary inspiral systems.  The horizontal axis in this plot represents the intrinsic chirp mass of a symmetric binary for the solid lines, and the observed chip mass for the dashed lines.  Blue lines represent the maximum observable distance for Advanced LIGO, whereas red lines represent the reach of Advanced LIGO with extended arms, based on the sensitivity shown in Figure~\ref{fig:sens}. A Hubble constant of 67.9~km/s/Mpc was assumed~\cite{ade:2013}. }
\label{fig:reach}
\end{center}
\end{figure}

A 40~km gravitational wave detector, with the sensitivity presented in Figure~\ref{fig:sens}, will so greatly change the distance at which sources can be observed that cosmological redshift must be accounted for when describing its potential reach. As for light, the expansion of the universe will shift gravitational wave signals down in frequency, moving signals from stellar mass objects into the most sensitive part of the band, and shifting signals from heavier sources below the detection band.

The frequency dependence of the expected waveforms for nearby compact object binaries is determined by the intrinsic chirp mass of the object,  $\mathcal{M}_0 = \sqrt[5]{\mu^3 M^2}$, with $\mu$ the reduced mass and $M$ the total mass.  The impact of a cosmological redshift on gravitational wave observations can be described entirely as a change in the observed chirp mass, $\mathcal{M}=(1+z)\mathcal{M}_0$, see Ref.~\cite{finn:1996}. The horizon distance for compact object binaries is defined as the maximum distance at which an optimally oriented system can be observed with a signal-to-noise ratio of 8; when the impact of cosmological redshift is negligible the horizon distance is about twice as far as the inspiral range which includes averaging over source orientation and sky location.  We plot the horizon distance as a function of intrinsic chirp mass in Figure~\ref{fig:reach}, as well as the horizon distance as a function of observed chirp mass.

As shown in Figure~\ref{fig:reach},  a pair of $1.4 M_\odot$ binary neutron stars, which has an intrinsic chirp mass of $1.2~M_\odot$ could be observed at a horizon redshift of about 2.  The observed chirp mass of this system, $\mathcal{M} \simeq 3.6 M_\odot$, can be found by looking at the intersection of the observed chirp mass curve with a line at $z = 2$. Note that since the signal from a binary system is redshifted into the detection band, the detector's reach for objects of this type is increased by about a factor of 2 in redshift.  On the other hand, the horizon distance for symmetric black hole systems with an intrinsic chirp mass above $\mathcal{M}_0 > 15 M_\odot$ is reduced by the cosmological redshift, since the waveform gets redshifted below the detection band.

With a 40~km observatory, the most distant detectable binary would have an intrinsic chirp mass of $\mathcal{M}_0\approx 5 M_\odot$ and a horizon redshift of $z=7.2$.  This means the reach extends into the latter part of the re-ionization epoch. While the rate of inspirals at these high redshifts will likely be low, observations of inspirals from the remnants of massive early stars may be possible, shedding light on the populations of early, metal poor stars.  Observations of binary black hole inspirals coupled with electromagnetic observations can provide a measurement of the distance-luminosity relation independent of the cosmological distance ladder, an important science goal for space based gravitational wave observatories like LISA \cite{Holz:2005}. Hence, coincident detections of high redshift sources would be able to provide measurements of cosmological parameters, including dark energy, which are completely independent of supernovae \cite{Sathyaprakash:2010}.

\section{Noise scaling with arm length}
\label{sec:scaling}

It would be easy to erroneously conclude that the sensitivity of a gravitational wave detector will scale linearly with increasing detector length, since the displacement caused by gravitational wave strain scales linearly with detector length.  However, \emph{all} of the limiting noise sources will also change as the detector length is changed, meaning that the sensitivity does not have a simple linear scaling with detector length at any frequency. Vertical motion of the optics, driven by suspension thermal noise, couples to the gravitational wave readout due to the curvature of the earth and does not scale linearly; coating thermal noise scaling is modified by the changing beam size; the mass of the optics must be increased to accommodate the larger beams; and the overall quantum noise behavior of the detector must be modified to account for the increased flight time of photons in the interferometer arms.

The power spectral density of the coating and substrate Brownian noise scales as the inverse of the laser beam area~\cite{levin:1998}. The spot sizes $w_1$ and $w_2$ on the mirrors in a two-mirror Fabry-Perot cavity are given by~\cite{siegman}
\begin{equation}
w_{1,2}^2 = \frac{\lambda L}{\pi} \sqrt{\frac{g_{2,1}}{g_{1,2}(1-g_{1,2} g_{2,1})}},
\label{equ:spotsize1}
\end{equation}
where $\lambda$ is the wavelength, $g_{1,2}=1-L_{\mathrm{arm}}/R_{1,2}$ are the $g$ factors for each optic, and $R_{1,2}$ are the radii of curvature of the two optics. The beam size on the optic scales with the square root of the arm cavity length if other factors are constant, meaning that the strain amplitude sensitivity limited by coating Brownian noise could improve as much as $1/L_{\mathrm{arm}}^{3/2}$ as the arm length increases, if suitably large optics are available. In reality, for a longer interferometer both the angular stability and the size of the required optics will require a smaller $g$ factor than Advanced LIGO, so that the scaling of Brownian noise will be between $1/L_{\mathrm{arm}}^{3/2}$ and $1/L_{\mathrm{arm}}$.  This increased beam size may require an increase in the mass of the optics used, which leads to a small improvement in the overall sensitivity due to reductions in the noise caused by Newtonian gravity, radiation pressure noise, and an even smaller reduction in the horizontal suspension thermal noise \cite{Saulson1990}.

Due to the curvature of the Earth, for multi-kilometer arm cavities the local vertical direction is not quite perpendicular to the optical axis, and this introduces a small but unavoidable coupling between vertical motion of the test mass and the gravitational wave output of the detector, approximately $\sin\left(L_{\mathrm{arm}}/2R_{\oplus}\right)$ (0.003 for a 40~km arm).  Even a small coupling can be problematic, because the vertical thermal noise is orders of magnitude larger than the noise in the horizontal direction, where a large fraction of the energy of oscillations is stored as gravitational potential energy.  In the vertical direction however, the energy is stored in the elastic restoring forces of the suspension fibers and springs, which introduce noise through their mechanical losses ~\cite{Gonzalez1994, Cagnoli2000}.  By lengthening the final suspension stage from 60 cm to 1 meter, the vertical suspension mode resonant frequency will be lowered from 9~Hz to 7~Hz~\cite{Hammond2012}. This modest change would reduce the suspension thermal noise by more than a factor of 7 at 10~Hz in a 40 km interferometer, while in a 4 km interferometer where the horizontal suspension noise dominates it would provide about a 30\% improvement.

Quantum noise is a combination of sensor noise (shot noise) and displacement noise (radiation pressure noise); the optical parameters of the interferometer must be chosen to optimize the quantum noise in light of the other limiting noise sources in the interferometer. At low frequencies the increased arm length improves the quantum noise limited sensitivity while at high frequencies the shot noise is unchanged as the arm length increases.  Quantum radiation pressure noise is reduced by the increased arm length because it is a displacement noise and because the fluctuating radiation pressure force causes smaller displacements in the more massive optics required for a longer detector.

At high frequencies the shot noise limited sensitivity does not change as the arm length increases, but can be improved by increasing the efficiency of the signal extraction~\cite{Mizuno:1993}, injection of squeezed light~\cite{Caves:1980,Caves:1981} and by increasing the circulating power. Since both thermal lensing and thermal distortion from heating of the optics due to absorption of laser light are approximately independent of the beam size~\cite{Winkler1991}, the circulating power in a long interferometer will be similar to that of Advanced LIGO. Since squeezed light injection is the most promising early upgrade for Advanced LIGO~\cite{oelker, Grote2011, LSCH1Sqz2013},  we assume that it will be included in any future interferometer designs. We include modest frequency-dependent squeezing with a 1~km long  filter cavity and 80~ppm round-trip losses~\cite{Evans2013,Isogai2013,Kwee2014}.

By increasing the efficiency of signal extraction, the detection band can be broadened by improving the shot noise limited sensitivity at high frequencies while slightly decreasing the quantum noise limited sensitivity from 30--80~Hz, where other noise sources also limit the sensitivity. Table~\ref{tbl:param} compares the optical parameters between Advanced LIGO and the 40~km extended version and includes the change in signal recycling mirror transmission required to maintain detection bandwidth.

\begin{table}[b]
\begin{tabular*}{3.4in}{@{\extracolsep{\fill}} l r r }
    \hline \hline
    & Adv. LIGO & 40\;km LIGO \\
    \hline
    Arm length & 4 km & 40 km \\
    Mirror mass & \multicolumn{2}{r}{40 kg\hspace*{0.4in}}\\
    Beam radius  & ~6.2 cm & 11.6 cm \\
    Measured squeezing & none & 5 dB \\
    Filter cavity length & none & 1 km\\
    Suspension length & 0.6 m & 1 m \\
    Signal recycling mirror trans. & 20\% & 10\%\\
    Arm cavity circulating power & \multicolumn{2}{r}{775 kW\hspace*{0.4in}}\\
    Arm cavity finesse & \multicolumn{2}{r}{446\hspace*{0.4in}}\\
    Total light storage time & 200~ms & 2~s \\
    \hline\hline
\end{tabular*}
\caption{Optical parameters of the Advanced LIGO detector and the 40~km extended version. The mirror mass may be increased in a larger interferometer to accommodate a larger beam size, leading to a slightly better sensitivity than that shown in Figure~\ref{fig:sens} due to reduction in radiation pressure noise.}
\label{tbl:param}
\end{table}

The statistical fluctuations in the column density of the residual gas in the vacuum system induce noise in the measured optical path of the laser beam~\cite{Zucker:1994}. These fluctuations are averaged over the entire length and size of the beam. For an ${\rm H}_2$ pressure of $5\times10^{-9}$ torr at room temperature, a level normally surpassed by the LIGO vacuum system, and the beam size listed in Table~\ref{tbl:param} the residual gas strain amplitude noise density is about $6\times 10^{-26}/\sqrt{\mathrm{Hz}}$, below the level of noise shown in Figure~\ref{fig:sens} and a factor of 5 below the limiting sensitivity.

\section{Constraints on arm length}

While many noise sources decrease with increasing arm length, there are several constraints which prevent indefinitely increasing the arm length. Cost is of course a huge consideration; here we consider two of the most important technical constraints: the laser spot size, which drives us to larger area optics and the increased challenges of maintaining interferometer alignment.

The first of these constraints arises from the necessary expansion of beam size with interferometer length due to diffraction, and the difficultly of manufacturing large optics with surfaces suitable for use in low-loss resonant cavities. For a spot of radius $w$, the clipping loss $p$ at a circular aperture (mirror) of radius r is given by
\begin{equation}
\log{ \left( p  \right)} =\frac{- 2 r^2}{w^2}.
\end{equation}
Advanced LIGO was designed for a total cavity round trip loss of $75~{\mathrm{ppm}}$, of which $1~{\mathrm{ppm}}$ per optic is clipping loss. If we allow an increase to $15~{\mathrm{ppm}}$ per optic for clipping and if we compensate with input laser power, we find for the maximum allowable arm length, with the simplification $g_1=g_2=g$:
\begin{equation}
L = \frac{2 \pi}{-\log{\left( p \right)} } \frac{r^2}{\lambda} \sqrt{1\!\!-\!\!g^2}
\approx 15\,{\mathrm{km}} \left( \frac{r}{17\,{\mathrm{cm}}} \right)^2 \!\! \sqrt{1\!\!-\!\!g^2},
\end{equation}
where we used Advanced LIGO's optics radius of $17~{\mathrm{cm}}$ and $\lambda=1.064~\mu\mathrm{m}$. With the goal of a ten-fold arm length increase over Advanced LIGO, this implies the need for optics with a diameter of about $55\,\mathrm{cm}$. This arguably is the toughest technical constraint to scaling up gravitational wave interferometers.

Optical surface quality requirements are driven by scattering losses in the arm cavities and contrast defect at the beamsplitter. The relevant spatial size of imperfections on the optics scales with the spot size $w$, i.e., it remains the same relative to the optic's diameter. Hence, the technical challenges of manufacturing suitable optics are not fundamental, but rather a question of adequate tooling and manufacturing capabilities. To keep the beam radius and, therefore, the optics small, lenses could potentially be used in the arm cavities. The noise requirement for such lenses is, however, stringent~\cite{RadiativeThermalNoise}.

The task of maintaining the interferometer alignment could be expected to become more challenging as the arm length is increased, especially during initial lock acquisition before active feedback servos can be engaged. Assuming a symmetric cavity ($g_1=g_2=g$) for simplicity, we find the loss due to a misalignment, $\theta_1$, of one of the mirrors to be proportional to cavity length
\begin{equation}
\label{eqn:loss}
P_{\mathrm{loss}} (\theta_1) =
\frac{\pi L}{\lambda} \frac{1}{(1-g^2)^{\frac{3}{2}}} \theta_1^2.
\end{equation}
To reduce coating Brownian noise by increasing the spot size, Advanced LIGO is already using a relatively high $g$-factor of $g^2=0.83$.  By choosing a smaller $g$-factor it is, therefore, possible to build a $40~{\rm km}$ arm cavity without enhancing the sensitivity to misalignment, and so existing suspension hardware may be sufficient even for a much longer interferometer.

Finding a suitable site for a 40~km long interferometer is challenging, but there are several relatively flat, undeveloped sites within the United States and around the world that could be suitable candidates. As examples we may list the Carson Sink in Nevada or the Murray river plane in Sedan, South Australia. Both sites are slight bowls, partly compensating for the Earth's curvature and therefore reducing the amount of earth moving needed. We expect that the disadvantages of location and cost for a long arm facility will be more than compensated for by the immense reduction in complexity and technical risk.

\section{Conclusion}

To summarize, a 40~km interferometer based on Advanced LIGO technologies, with modest levels of squeezed light injection and the minimum beam size possible without focusing optics, can be made an order of magnitude more sensitive than Advanced LIGO. We emphasize again that a factor of 10 change in length doesn't necessarily result in a factor of ten change in sensitivity; the modifications in optical parameters detailed in Table~\ref{tbl:param} were carefully chosen to make this possible.

While the advantages of scaling up current interferometers have some limitations, a factor of 10 scaling is nearly optimal, as it enables the detection of astrophysical events from much of the visible universe. A typical $1.4 M_{\odot}$ binary neutron star system can be detected at a redshift of $z \sim 2$, and a symmetric $10 M_\odot$ black hole binary can be detected back to the epoch of reionization at $z\gtrsim 7$. 
The detector described herein will do more than provide more frequent detections, it will open up new scientific possibilities for gravitational wave astrophysics. High signal to noise observations of the sources, accessible to current detectors only at modest fidelity, will allow studies of gravity in the strong field dynamical regime; and it will better reveal the properties of the compact objects involved (e.g., the neutron star equation of state). The reach of this detector will include a significant part of the history of star formation and allow observation of most solar mass compact object binary inspirals throughout the universe. Increased sensitivity will also bring observations of sources rare or unseen by current detectors, such as supernovae and continuous wave signals from spinning neutron stars.

Furthermore, the investment in a new 40~km facility provides the opportunity to integrate more advanced technologies in the future---limited only by fundamentals like the speed of light and the curvature of the Earth.

\begin{acknowledgments}
We thank Rai~Weiss for encouraging us and for thinking about ways to reduce the costs of a 40~km interferometer. We would like to thank Kiwamu Izumi, David Ottaway, GariLynn Billingsley, and Stefan Hild for many fruitful discussions. This work was supported by the National Science Foundation grants PHY-0823459 and PHY-1068809. This document has been assigned the LIGO Laboratory document number LIGO-P1400147.
\end{acknowledgments}

\bibliographystyle{apsrev4-1}
\bibliography{LongIfoScaling}

\end{document}